\documentstyle[11pt,epsf]{article}
\input psfig
\def\beq{\begin{eqnarray}}
\def\eeq{\end{eqnarray}}
\def\bea{\begin{eqnarray*}}
\def\eea{\end{eqnarray*}}

\def\centeron#1#2{{\setbox0=\hbox{#1}\setbox1=\hbox{#2}\ifdim
\wd1>\wd0\kern.5\wd1\kern-.5\wd0\fi
\copy0\kern-.5\wd0\kern-.5\wd1\copy1\ifdim\wd0>\wd1
\kern.5\wd0\kern-.5\wd1\fi}}
\def\ltap{\;\centeron{\raise.35ex\hbox{$<$}}{\lower.65ex\hbox{$\sim$}}\;}
\def\gtap{\;\centeron{\raise.35ex\hbox{$>$}}{\lower.65ex\hbox{$\sim$}}\;}

\def\singleandthirdspaced{\baselineskip=\normalbaselineskip\multiply
    \baselineskip by 130\divide\baselineskip by 100}

\newcommand{\newc}{\newcommand}
\newc{\qbar}{{\overline q}}
\newc{\Kahler}{K\"ahler }
\newc{\deltaGS}{\delta_{\rm GS}}
\begin{document}
\begin{titlepage}
\begin{flushright}
{\large hep-th/0506246 \\ SCIPP-2005/02\\
}
\end{flushright}

\vskip 1.2cm

\begin{center}

{\LARGE\bf R Symmetries in the Landscape}

\vskip 1.4cm

{\large  M. Dine and Z. Sun}
\\
\vskip 0.4cm
{\it Santa Cruz Institute for Particle Physics,
     Santa Cruz CA 95064  } \\

\vskip 4pt

\vskip 1.5cm

\begin{abstract}
In the landscape, states with $R$ symmetries at the classical level form a distinct
branch, with a potentially interesting phenomenology.  Some preliminary analyses
suggested that the population of these states would be significantly
suppressed.  We survey orientifolds of IIB theories compactified
on Calabi-Yau spaces based on vanishing polynomials in
weighted projective spaces, and find that the suppression
is quite substantial.  On the other hand, we find that
a $Z_2$ $R$-parity is a common feature in the landscape.
We discuss whether
the cosmological constant and proton decay or
cosmology might select the low energy branch.
We include also some remarks on split supersymmetry.
\end{abstract}

\end{center}

\vskip 1.0 cm

\end{titlepage}
\setcounter{footnote}{0} \setcounter{page}{2}
\setcounter{section}{0} \setcounter{subsection}{0}
\setcounter{subsubsection}{0}

\singleandthirdspaced

\section{Introduction}

Recent studies of string configurations with fluxes have provided support
for the idea that string theory possesses a vast landscape of string\cite{landscape1,
landscape2,landscape3,bdg,landskepticism}.
In the landscape, three distinct branches of states have been identified\cite{dgt}.
One branch has broken supersymmetry already in the leading approximation.
Another has unbroken supersymmetry at tree level, with negative cosmological
constant.  A third has unbroken supersymmetry and vanishing cosmological
constant at tree level.  Non-perturbatively, we might expect that supersymmetry
breaking occurs generically in the latter two cases, so the distinctions between
these states, individually, are not sharp.  However, the statistics of these three
branches are quite distinct.  The first, ``non-supersymmetric branch" has a distribution
of states which strongly peaks at the highest energy scale; states with a low scale
of supersymmetry breaking, $m_{3/2}$, are suppressed by $m_{3/2}^{12}$\cite{douglasdenef2,
dos}.  The
second, ``intermediate scale branch", has a distribution of scales roughly logarithmic
in $m_{3/2}$, $\int{dm_{3/2}^2 \over m_{3/2}^2 \ln(m_{3/2})}$.  The third, ``low scale" branch 
will be the focus of this paper.  Here the distribution behaves as
\beq
\int {dm_{3/2}^2 \over m_{3/2}^4}
\eeq

For unbroken supersymmetry, vanishing of the cosmological constant implies the
vanishing of $W$.  
Vanishing of $W$ is often connected with $R$
symmetries. $R$ symmetries
are symmetries which transform the supercharges non-trivially.  Among these,
we can consider two broad classes, those which transform the superpotential
by a non-trivial phase, and $Z_2$ symmetries under which
the superpotential is invariant.  Conventional $R$-parity is
an example of the latter, and we will refer
to such $Z_2$ symmetries more generally as $R$-parities.  These are not, in any
general way, connected with vanishing $W$.  We will reserve the term
$R$ symmetry for those symmetries which transform $W$.
In \cite{dgt} it was argued that states on the low scale branch were likely to arise
as a result of discrete $R$ symmetries. 
In \cite{dos}, some aspects of these states
were considered and some counting
performed.  In typical constructions of flux vacua, an $R$ symmetry
can arise if the underlying theory, in the absence of fluxes, possesses such
a symmetry, and if the non-vanishing fluxes are themselves neutral under the symmetry.
As explained in \cite{dos}, for a reasonably generic superpotential consistent with
the symmetries, the potential
has stationary
points preserving
both supersymmetry and $R$ symmetry\footnote{In \cite{DeWolfe:2004ns}
and \cite{dewolfe}, explicit features of the superpotential\cite{gvw} are employed to
actually find and count solutions with these properties}.
This can be illustrated by compactification of the IIB theory on an orientifold of the
familiar quintic in $CP^4$.  On a subspace of the moduli space, prior
to performing the orientifold projection, the quintic is known to
possess a large discrete symmetry, $Z_5^4 \times S_5$\cite{newissues,gsw}.  The
projection can preserve a subgroup of this group.  It is not difficult
to classify the fluxes according to their transformation properties under these
symmetries\footnote{We will correct an error in the identification of symmetries
in \cite{dos}, but this will not qualitatively alter the earlier conclusions}.
If one tries to turn on fluxes
in such a way as to preserve a single $Z_5$, one finds that it is necessary to set more
than 2/3 of the fluxes to zero.  In landscape terms, this means that the dimensionality of
the flux lattice is reduced by 2/3, and correspondingly there is a drastic reduction
in the number of states.  One of the principle goals of the present paper is to assess
whether this sort of reduction is typical.

One of the observations of \cite{dos} is that in the bulk of $R$-symmetric states,
supersymmetry and $R$ symmetry are likely to be unbroken.  We will explain this
observation further,
and discuss the assumptions on which it relies.

Discrete symmetries are of interest for other reasons.  One of the most important is to
suppress proton decay.  Usually one considers $R$ parity, but more general $R$ symmetries
can suppress not only dimension four but also dimension five operators.  $R$ parity
is distinctive in that it does not rotate the superpotential.  As a result, it need not
be spontaneously broken (it does not forbid a mass for gauginos).  It also does not lead
to a non-vanishing $\langle W \rangle$.  So states on the intermediate branch can be
$R$-parity-symmetric. We will discuss the distinctions between $R$ parity and $R$
symmetries further in this paper.

$R$ symmetries have received attention recently for another reason:  they are
part of the rationale for the ``split supersymmetry" scenario\cite{Arkani-Hamed:2004fb}.
We will discuss a number of issues related to this proposal here as well.

This paper is organized as follows.  In the next section we will describe a counting
exercise based on Calabi-Yau models constructed as complete intersections in
weighted projective spaces.  We will see that the results for the quintic are rather
general:  we find no examples where more than $1/3$ of the possible fluxes
preserve an $R$ symmetry.  In section 3, we verify our identification of discrete
symmetries of these models by studying Gepner models\cite{gepner}.
In sections four and
five, we explain why supersymmetry and $R$ symmetry are typically unbroken
in these states at the classical level,
and consider non-perturbative effects which
can break these symmetries.  In the final sections,
we discuss split supersymmetry
and $R$ parity.  We explain why split supersymmetry
seems an unlikely outcome of the
landscape and contrast $R$-parity and more general $R$ symmetries.  We conclude
with a discussion of
selection effects which might favor one or another branch of the landscape.

\section{R Symmetries in Weighted Projective Spaces}

Already in \cite{newissues,gsw}, the existence of discrete symmetries in Calabi-Yau
spaces has been noted.  It is instructive to
enumerate these symmetries before implementing
the orientifold projection.  These symmetries can be thought of as discrete subgroups
of the original Lorentz invariance of the higher dimensional space.
The quintic in $CP^4$ provides a familiar
example.
The construction of the Clabi-Yau space begins with a choice of a vanishing
quintic polynomial.
The polynomial
\beq
P= \sum_{i=1}^5 z_i^5 = 0
\eeq
exhibits a large discrete symmetry.  Each of the $z_i$'s can be multiplied by 
$\alpha= e^{2 \pi i \over 5}.$  In addition, there is a permutation symmetry
which exchanges the $z_i$'s.  To see that these are $R$ symmetries, one can proceed
in various ways.  One can, first, construct the holomorphic three form.  Defining
variables $x_i = z_i/z_5$, this can
be taken to be\cite{gsw}:
\beq
\Omega = dx_1 dx_2 dx_3 \left ( {\partial P \over \partial x_4}\right )^{-1}.
\label{omegaequation}
\eeq
It is easy to check that as long as ${dP \over dz_i} \ne 0, i=1,\dots,5$ (the
transversality condition) this formula is ``democratic"; the singling out of $z_4$
and $z_5$ is not important.  $\Omega$ transforms under any symmetry like the superpotential.
This follows from the fact that $\Omega_{IJK} =  \eta^T \Gamma_{IJK} \eta$,
where $\eta$ is the covariantly constant spinor.
So we can read off immediately that under, say, $z_1 \rightarrow \alpha z_1$,
the superpotential transforms as $W \rightarrow \alpha W$.  Similarly, under
an odd permutation, the superpotential is odd.

The complex structure moduli are in one to one correspondence with deformations
of the polynomial $P$, so it is easy to determine their transformation properties
under the discrete symmetry.  Overall, there are $101$ independent polynomials.
So, for example, the polynomial $z_1^3 z_2^2$ transforms as $\alpha^3$ under
the symmetry above.  $z_1^4 z_5$, on the other hand,
is not an independent deformation, since it can be absorbed
in a holomorphic redefinition of the $z_i$'s.
For the landscape, it is also important to understand how
the possible fluxes transform:  fluxes are paired with complex structure moduli.
Because they correspond to $RR$ states, they transform differently than the
scalar components of the moduli.  As we explain below, the criterion that a flux
not break and $R$ symmetry is that {\it the corresponding modulus transform
under the symmetry like the holomorphic $3$ form.}  
The effective lagrangian for the light fields will exhibit a symmetry if all
fluxes which transform non-trivially vanish.

To see how the transformation properties of the fluxes relate to those of the
moduli, we can proceed by using equation \ref{omegaequation}
to construct the holomorphic three form.
If we deform the polynomial by
$P \rightarrow P + \psi~ h(z_i)$, then:
\beq
\delta \Omega = dx^1 \wedge dx^2 \wedge dx^3~ \psi~{\partial h \over \partial x^4}
({\partial P \over \partial x^4})^{-2}.
\eeq
If this is to be invariant, the transformation of $h$ must compensate
that of $dx_1 \dots dx_4$  Since $h$ transforms like $\psi$, we see that $\psi$ must
transform like $\Omega$.

So in the case of the quintic, consider the transformation $z_1 \rightarrow \alpha
z_1$, where $\alpha= e^{2 \pi i
\over 5}$.  Under this transformation, $\Omega$ transforms like $\alpha$.  So the
invariant fluxes correspond to polynomials with a single $z_1$ factor.  Examples
include $z_1 z_2^2 z_3$, $z_1 z_2^2 z_3^2$ and $z_1 z_2 z_3 z_4 z_5$. Altogether, of
the $101$ independent polynomial deformations, $31$ transform properly.

However, we need to consider the orientifold projection.  In the IIB theory,
this projection takes the form\cite{grimmlouis}:
\beq
{\cal O} = (-1)^{F_L} \Omega_p \sigma^*~~~~~~\sigma^* \Omega = -\Omega.
\eeq
Here $\Omega_p$ is orientation reversal on the world sheet; $\sigma$ is a space-time
symmetry transformation.  In the case of the quintic, a suitable $Z_2$ transformation
can be found among the various permutations.  An example is the cyclic transformation:
\beq
z_2 \rightarrow z_3 ~~~ z_3 \rightarrow z_4  ~~~
z_4 \rightarrow z_5~~~z_5 \rightarrow z_2.
\eeq
There are $27$ polynomials invariant under this
symmetry, so
$h_{2,1}$
is reduced from $101$ to $27$.  The number of fluxes which are invariant
under the symmetry is reduced to $9$.  This is only $1/3$ of the
total.

In the flux landscape, it is the fact that there are a large number of possible
fluxes which accounts for the vast number of states. If one thinks of the fluxes
as forming a spherical lattice, it is the large radius of the
sphere and the large dimension of the space which account for the huge
number of states.  Reducing the
dimensionality significantly drastically reduces the number
of states; e.g. if 2/3 of the fluxes must be set to zero,
$10^{300}$ states becomes $10^{100}$.   In the case of the quintic, we
we have just seen that requiring,
for example, the $z_1 \rightarrow \alpha z_1$ symmetry requires that more
than 1/3 of the fluxes
vanish.  In the end, though, the dimension of the flux lattice was not
so large in this case.
A natural question is whether such a large fractional reduction
in the dimensionality of the lattice is typical.

A large class of Calabi-Yau spaces have been constructed as hypersurfaces in weighted
projective spaces\cite{cyweb}.  The corresponding polynomials can exhibit complicated sets of discrete
symmetries.
  Here we will consider some
examples chosen from the list.  

A case in which there is a large number of fluxes even after the orientifold
projection is
provided by $WCP^4_{1,1,1,6,9}[18]$.  This model is, for a particular
radius and choice of polynomial, one of the Gepner models\cite{gepner}
and so we have more than one check on our analysis.  Take the polynomial to be:
\beq
P=z_1^{18} + z_2^{18} + z_3^{18} + z_4^3 + z_5^2 =0.
\label{polynomialb}
\eeq
Then there are $h_{2,1}= 272$ independent deformations of the polynomial.
There is also a rich set of discrete symmetries:
\beq
Z_{18}^3 \times Z_3 \times Z_2 \times S_3.
\eeq
One can construct $\Omega$ as in eqn. \ref{omegaequation}; one finds that under
$z_1 \rightarrow e^{2 \pi i \over 18} z_1$,
$\Omega$ transforms as:
\beq
\Omega \rightarrow e^{2 \pi i \over 18} \Omega,
\eeq
and similarly for the other coordinates.  In particular, $\Omega \rightarrow -\Omega$
under the transformation $z_5 \rightarrow -z_5$.
Now all of the polynomials are invariant under the $Z_5$.
Any polynomial linear in $z_5$ can be absorbed into a redefinition of $z_5$
(just as the $z_i^4 z_j$ type polynomials to not correspond to physical deformations
in the case of the quintic).  So all of the fluxes are odd under the symmetry.
So if we take this to be the $\sigma$ of the orientifold projection,
then all of the complex structure moduli and the fluxes survive.

Now we want to ask:  what fraction of the fluxes preserve
a discrete symmetry of the orientifold theory.
Consider, for example, $z_4 \rightarrow e^{2 \pi i \over 3}
z_4$.  Invariant fluxes are paired with polynomial deformations linear in $z_4$.  There
are $55$ such polynomials.  So, as in the case of the quintic,
approximately $1/3$ of the fluxes
are invariant under the symmetry.  Indeed, surveying numerous models and many symmetries,
we have found no examples in which 1/2 or more of the fluxes are invariant.
The model $WCP^4_{1,1,1,6,9}[18]$
is particularly interesting, since it has the largest $h_{2,1}$
in this class.

In the next section, to confirm our identification of these symmetries, we discuss
$R$ symmetries in the Gepner models.  

\section{Identifying $R$ Symmetries in the Gepner Models}

A number of the models in weighted projective spaces have
realizations as Gepner models\cite{gepner}.  These provide a useful laboratory
to check our identification of symmetries and field transformation
properties.
We adopt the notation of \cite{distlergreene}.  States of the full theory are products
of states of $N=2$ minimal models with level P.  These are labelled:
\beq
\left ( \ell ~\matrix{ q & s \cr \bar q & \bar s} \right )
\eeq
Here $\ell = 0,\dots, P$, and $\ell +q+s =0 ~{\rm mod}~ 2$.
The right-moving conformal weight and $U(1)$ charge are:
\beq
h= {{1 \over 4} \ell (\ell+2)-q^2/4 \over (P+2) }
+ {1 \over 8} s^2; ~~ Q = {-q \over P+2} + {1 \over 2} s.
\eeq
and similarly for the left movers.
Each of the minimal models has a $Z_{P+2}$ symmetry; states transform
with a phase:
\beq
e^{-{i \pi (q + \bar q) \over P+2}} 
\eeq
The right-moving supersymmetry operator is a product of operators in each of the minimal models of the form:
\beq
S=\left ( 0 ~\matrix{ 1 & 1 \cr  \bar 0 & \bar 0} \right ).
\eeq
From this we can immediately read off the transformation properties of $S$ under the
discrete symmetries, and determine whether or not the symmetries are $R$ symmetries.

The quintic in $CP^4$ is described by a product of $5$ models with $P=3$.  Following Gepner,
we can identify the complex structure moduli associated with various deformations of
the symmetric polynomial by considering their transformation properties under the
discrete symmetries.  So, for example, the polynomial $z_1^3 z_2$ is identified with
the state:
\beq
\left ( 3 ~\matrix{ 3 & 0 \cr 3 & 0} \right )
\left ( 2 ~\matrix{ 2 & 0 \cr 2 & 0} \right )
\left ( 0 ~\matrix{ 0 & 0 \cr 0 & 0} \right )^3.
\eeq
One can enumerate all of the states in this way, and repeat
the counting we did before.

Now consider the model $WCP^4_{1,1,1,6,9}[18]$, with the polynomial of equation
\ref{polynomialb}.  For a particular choice of radius, this is described by
the Gepner model which is the product $(16,16,16,1)$.  We see that the
symmetry is $Z_{18} \times Z_{18} \times Z_{18}
\times Z_3$.  The $Z_2$ which takes the coordinate
$Z_5$ of the weighted projective space into minus itself, $z_5 \rightarrow -z_5$,
is equivalent, because of the identifications
of the weighted projective space, to the transformation:
\beq
z_{1,2,3} \rightarrow e^{2 \pi i \over 18} z_{1,2,3} ~~~~~ z_4 \rightarrow e^{2 \pi i \over 3}
z_4.
\eeq
This is an $R$ symmetry; it multiplies $S_\alpha^2$, and hence the superpotential,
by $-1$.  Again we can enumerate
the states.  For example, the polynomial $z_1^{16} z_2^{2}$ is identified with the
operator:
\beq
\left ( 16 ~\matrix{ 16 & 0 \cr 16 & 0} \right )
\left ( 2 ~\matrix{ 2 & 0 \cr 2 & 0} \right )
\left ( 0 ~\matrix{ 0 & 0 \cr 0 & 0} \right )^3.  
\eeq
This is clearly invariant under the symmetry above.  It is a simple matter to enumerate all
of the possible states and check that they are invariant.

So, as we stated earlier, all of the fluxes are invariant under the $Z_2$,
since the scalar moduli are odd.
It is a simple matter to check that the supercharges transform
by $-1$ under the $R$ parity symmetry we identified earlier, and to
reproduce our counting for the $Z_{18}$ symmetry as well.

It is particularly easy to survey models which have realizations as Gepner models.
We have found no examples where more than $1/3$ of the fluxes are invariant
under an $R$ symmetry.

\section{Supersymmetry and R-Symmetry Breaking at Tree Level}

One can ask whether in a theory with $R$ symmetries, supersymmetry and $R$ symmetry
are spontaneously broken.  We have seen that invariant fluxes are paired with
moduli which transform like the superpotential.  We have also seen that typically
less than $1/2$ of the moduli transform in these way.

Call $X_i$, $i=1,\dots N$, those moduli which transform like the superpotential under
$R$ symmetries. Denoting the other fields by $Y_a$,
These break into two groups: those invariant under the $R$ symmetry, $\phi_\alpha$,
$\alpha=1,\dots P$,
and those which transform in some way, $\chi_r$.  Including terms at most linear in
fields which transform under the $R$ symmetry, 
the superpotential has the form:
\beq
W = \sum_{i=1}^N X_i f_i(\phi_\alpha)
\eeq
If $N \le P$, then
provided that the $f_i$'s are reasonably generic functions, the equations
$f_i=0$ have solutions, so there are vacua with $X_i=f_i=0$, and supersymmetry
and the $R$ symmetry are unbroken.  Consider our example based on   
$WCP^4_{1,1,1,6,9}[18]$.  We studied there the $Z_3$ symmetry, $z_4 \rightarrow
e^{2 \pi i \over 3} z_4$, and saw that there are $55$ $X_i$ fields, i.e.
N=55..  There are 
many more $\phi$ fields (corresponding to polynomials with no $z_4$); $P= 217$.
So among the fluxes which are invariant under the symmetry, generically one
expects to find supersymmetric, $R$ symmetric stationary points of the action.
Another symmetry one can study is the symmetry $z_1 \rightarrow e^{2 \pi i \over 18}
z_1$.  There are $28$ fluxes invariant under the symmetry, and correspondingly 
$N=28$ (these are all of the polynomials linear in $z_1$).  It turns out that there
are $28$ polynomials invariant under the symmetry.  So in this case, $N=P$,
and again one expects $R$-symmetric, supersymmetric solutions.
What is striking is that if one does find vacua with $N$ close to $h_{2,1}$,
so that there might not
be a huge suppression, {\it supersymmetry and/or $R$ symmetry typically will be broken}.
This situation, if it occurs, might be relevant to the ideas
of split supersymmetry, which we discuss further below.

\section{Non-Perturbative Mechanisms for Supersymmetry and R-symmetry Breaking}

We have seen that discrete $R$ symmetries tend
to give solutions
of the classical equations with unbroken supersymmetry and vanishing
$W$ (and hence vanishing cosmological constant) with very mild assumptions
about the form of the superpotential.  
Even if supersymmetry and $R$ symmetry are unbroken at the level of the classical
analysis, one expects that generically they will be broken by quantum effects.
We can speculate on a number of breaking mechanisms.  First, discrete symmetries
may suffer from non-perturbative anomalies\cite{discreteanomalies}.
As a result, non-perturbative effects
can generate an explicit violation of the symmetry.  In generic states in the landscape,
the couplings are presumably strong, so there is no real sense in which the
theory possesses such a symmetry at all.  But in a significant subset, these effects may
be small (e.g. exponential in small couplings).  Such effects could, in addition
to breaking the $R$ symmetry, break supersymmetry and generate positive and negative
contributions to the cosmological constant.

Another possibility is that the $R$ symmetry might be broken spontaneously by low
energy dynamics.\footnote{Because the discrete symmetries are gauge symmetries, the
distinction between explicit and spontaneous breaking has limited meaning, but the
terminology is useful here nevertheless.}  Gaugino condensation is an obvious example,
which spontaneously breaks any $R$ symmetry.  Models of dynamical supersymmetry
breaking generically break $R$ symmetries as well.

All of these effects are typically exponentially small as some coupling goes to
zero.  As a result, the scales of supersymmetry and $R$ symmetry breaking tend
to be distributed roughly uniformly on a log scale.  This was the basis of the
argument of \cite{dgt} for the distribution of states on this branch.

\section{Observations on Split Supersymmetry}

The authors of \cite{Arkani-Hamed:2004fb} made the interesting observation that if one simply
removes the squarks and sleptons from the MSSM, coupling unification works
as well or better than if these fields are at the TeV scale.  They suggested
that such a splitting of the spectrum might be typical of the landscape.
For example, we are used to the idea that fermion masses are often protected
by chiral symmetries, while something like supersymmetry is required to protect
scalar masses.

Upon further thought, however, there is a problem with this idea.  The fermions
whose masses one wants to protect are the gauginos.  In $N=1$ theories,
the only symmetries which can protect gaugino masses are $R$ symmetries.
But in the context of supergravity, if supersymmetry
breaking is large and the cosmological constant is small,
the $R$ symmetry is necessarily badly broken by the non-vanishing expectation
value of the superpotential.  At best, one expects that gaugino masses will be
suppressed relative to squark masses by a loop factor.  Ref. \cite{Arkani-Hamed:2004fb}
constructed field theory
models with larger suppression, but it is not clear that the features of these
models are typical of regions of the landscape.
One could speculate that there might be some anthropic selection
for a dark matter particle, but this would at best explain why one gaugino was
tuned to be light, not the three required for successful unification.

We have seen, in addition, that the studies of IIB vacua suggest that in
the bulk of $R$-symmetric states, supersymmetry and $R$ symmetry are likely
to be unbroken at tree level, and the statistics of these states suggests that
the vast majority of states with small cosmological constant will have small
supersymmetry and $R$ symmetry breaking.  One can legitimately object
that the IIB states might not be suitably representative.  In particular,
this argument relies crucially on a pairing of moduli and fluxes, which
might not hold in all regions of the landscape.

%

Suppose we do find $R$ symmetric flux configurations
for which the superpotential does not have $R$-symmetric, supersymmetric
stationary points (i.e. configurations for which there are more $X$-type
than $\phi$-type moduli).  Let's ask how
natural it might be to preserve the $R$ symmetry if supersymmetry
is broken.  Usually, preservation of a symmetry
is technically natural, since it is simply a question
of a sign of a particular term in an effective action.
In the case of the landscape, however, where there
are many fields, preserving a symmetry requires that
{\it many terms} in the action have the same
sign.  The authors of\cite{Arkani-Hamed:2004fb}
discuss this issue in some toy models, where perturbative corrections all
have the same sign.  In the framework of supergravity models, already at
the classical (tree) level, potentials for the moduli appear, and one
can ask what happens.  We have not performed a general analysis, but as a
toy model, have considered the $T_6/Z_2$ orientifold, where the Kahler 
potential can be written explicitly. With various assumptions about
supersymmetry breaking, one typically finds that at stationary
points of the potential with unbroken $R$ symmetry, some moduli transforming
under the symmetry have negative masses, some positive masses.

In any case, in order to understand the smallness of the cosmological constant,
at least within any semiclassical analysis, it is necessary that the $R$ symmetry
be very badly broken so that $\langle W \rangle$ is large.

\section{R Parity}

We have seen that $R$ symmetries are quite costly in the landscape.  Only a tiny
fraction of states in the flux vacua respect any $R$ symmetry.  $R$ parity is different,
however.  In many cases, there is an $R$ parity which is respected by
all of the fluxes.  Consider, again,
$WCP^4_{1,1,1,6,9}[18]$.  We study the
$Z_2$ symmetry:
\beq
z_i \rightarrow e^{4 \pi i \over 9} z_i, i=1\dots3; z_4 \rightarrow e^{2 \pi i \over 3}
z_4.
\eeq 
Under this symmetry, $\Omega$ is invariant, but the supercharges transform with a $-1$
(this is clear from our formulas for the Gepner version of the 
model).  Because $\Omega$ is invariant, fluxes are invariant if the corresponding
polynomial is invariant.  It is easy to check that every polynomial is invariant
under the $Z_2$.  (These symmetry properties are readily checked in the Gepner
construction as well).  This sort of symmetry appears in many of the models.

Such $Z_2$ R parity does not lead to $W=0$ vacua.  The typical state
in this case lies on the intermediate scale or high
scale branch of the landscape ($W \ne 0$, supersymmetry broken or unbroken).
So $R$ parity is a common feature of the 
intermediate scale branch of the landscape.

\section{Conclusions:  Phenomenology on the Low Energy Branch of the Landscape}

It would be exciting if one could argue that the low energy branch of the landscape
were favored.  This branch is likely to have a phenomenology similar to that of 
gauge-mediated models.  However,
we have seen that $R$ symmetry is rather rare in the landscape, even as $R$ parity
is common.   We can ask whether there are effects which might select for the
low energy branch.  Possibilities include:
\begin{enumerate}
\item
The cosmological constant:  on the low energy branch, very low scales for supersymmetry
breaking are favored.  So many fewer states are required than on the other branches to
obtain a suitably small cosmological constant.  If one supposes that the supersymmetry
breaking scale is, say, $10$ TeV, while that on the intermediate scale branch is
$10^{11}$ GeV, one needs $10^{28}$ fewer states.  But our analysis here suggestions that
the suppression of states on the low energy branch is far larger.
\item
Proton decay:  $R$ symmetries can account for the absence of proton decay.  But
we have seen $R$ parity is much more common than $R$ parity, so the latter would seem
a more plausible resolution to the problem of proton decay.  
\item
Cosmology:  The low energy branch has a severe cosmological moduli problem.  If SUSY
is broken at, say, $100$ TeV, the moduli are extremely light and dominate the energy
density of the universe at very early times.
This leads to too much dark matter, and it is likely that this is selected against.
\end{enumerate}

All of these considerations strongly suggest that the low energy branch of the
landscape is disfavored.  We have given elsewhere arguments which might favor
the intermediate scale branch, and explored its phenomenology\cite{intermediate}.  The recognition
that $R$ parity is common provides further support for this branch.

\noindent
{\bf Acknowledgements:}
We thank S. Kachru and O. DeWolfe for conversations and patient explanations, which
more than once kept us from going off on the wrong track.  We particularly
thank O. DeWolfe for communicating some of his results to us before publication.
T. Banks, D. O'Neil and S. Thomas also made valuable suggestions.
This work supported in part by the U.S.
Department of Energy.

\end{document}